\documentclass[twocolumn,showpacs,preprintnumbers,prl,aps]{revtex4}
\newcommand{\BE}{\begin{equation}}
\newcommand{\EE}{\end{equation}}
\newcommand{\BA}{\begin{eqnarray}}
\newcommand{\EA}{\end{eqnarray}}

\input epsf.tex

\begin{document}

\title{Bandgaps in the propagation and scattering of surface water waves over cylindrical steps}
\author{Liang-Shan Chen$^1$} \author{Chao-Hsien Kuo$^2$} \author{Zhen
Ye$^2$}\email{zhen@phy.ncu.edu.tw}\author{Xin Sun$^{1,3}$}
\affiliation{$^1$Department of Physics, Fudan University,
Shanghai 200433, P. R. China\\
$^2$Wave Phenomena Lab, Department of Physics, National Central
University, Chungli, Taiwan 32054, R. O. China\\ $^3$National Lab
of Infrared Physics, Shanghai Institute of Technical Physics,
Shanghai 200083, P. R. China}
\date{\today}

\begin{abstract}

Here we investigate the propagation and scattering of surface
water waves by arrays of bottom-mounted cylindrical steps. Both
periodic and random arrangements of the steps are considered. The
wave transmission through the arrays is computed using the
multiple scattering method based upon a recently derived
formulation. For the periodic case, the results are compared to
the band structure calculation. We demonstrate that complete band
gaps can be obtained in such a system. Furthermore, we show that
the randomization of the location of the steps can significantly
reduce the transmission of water waves. Comparison with other
systems is also discussed.

\end{abstract}


\maketitle

Phenomena pertinent to waves in complex media have been and
continue to be a great inspiration for the development of
scientific explorations. One of the most important phenomena is
frequency band structures. It prevails when waves propagate
through periodically structured media. Such a phenomenon was first
investigated for electrons in solids nearly eighty years ago. The
well-known Bloch theorem has then been proposed and led to the
successful explanation of such important properties of solids as
conductivity, semi-conductivity, and insulating
states\cite{Mermin}. Applying these concepts to classical
waves\cite{Bri,Yab} has paved an avenue to the new era of
research. Not only all the phenomena previously observed or
discussed only for electronic systems are successfully
transplanted to classical systems, but many more significant and
novel ideas and applications, such as photonic-crystal-made
negatively-refractive devices\cite{NR}, have well gone beyond
expectation, and are so far reaching that a fruitful new field has
been established, {\it i.~e.} the field of photonic and acoustic
crystals, signified by the establishment of a comprehensive
archive website\cite{web}.

Recently, the consideration of waves in periodic media has also
been deliberately extended to the propagation of water waves over
periodically structured bottoms\cite{Hare,Chou,Porter,Nature,APL}.
Some of the advances have been reviewed, for example, by
McIver\cite{McIver}. One of the most recent pioneering experiments
used water waves to illustrate the phenomenon of Bloch waves as a
result of the modulation by periodic bottom
structures\cite{Nature}. This experiment made it possible that the
abstract concept be presented in an unprecedentedly clear manner.

Motivated by the experiments\cite{Nature}, in this paper we would
like to further explore the propagation of water waves through
underwater structures. The structures considered here consist of
arrays of cylindrical steps mounted on a flat bottom. Although
there have been many theoretical approaches for investigating
propagation of water waves over various bottom topographies, most
approaches have been limited to mildly varying bottom structures
or their variations\cite{CCM,Ding}. In this paper, we will use the
recently derived theory \cite{Ye}. The main reason lies in that
the theory was successfully in explaining the experimental
observations\cite{PRE}. In addition, it has been shown that this
approach compares favorably with existing approximations when
applied to some special cases considered previously. We will
calculate the wave transmission and band structures for
periodically arranged arrays. Then we will show the effect of
positional disorders on the transmission. The results suggest that
the phenomenon of complete band gaps by analogy with the photonic
crystals is also possible for water waves. The complete band gaps
refer to the frequency regime where waves cannot propagate in any
direction.

Before continuing, it is worth noting here that from a more
general perspective, propagation of water waves over topographical
bottoms has been a subject of much research from both practical
and theoretical aspects since Lamb\cite{Lamb}. From the practical
side, the topic is essential to many important ocean engineering
problems such as floating bridges and devices in offshore power
stations\cite{McIver} A great amount of papers and monographs has
been
published\cite{Long,Bar,Miles,Peregrine,Meyer,Yue,JFM,Evans,Newman}.
A comprehensive reference on the topic can be found in two
excellent textbooks\cite{CCM,Ding}.

\begin{figure}[hbt]
\vspace{10pt} \epsfxsize=3in\epsffile{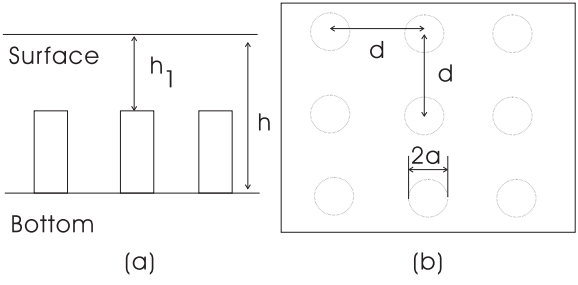}
\caption{Conceptual layout of the system: (a) side view; (b)
bird's view. Here the cylindrical steps with height $\Delta h = h
- h_1$ form a rectangular array with lattice constant $d$.}
\label{fig1}
\end{figure}

A conceptual layout of the system considered here is presented in
Fig.~\ref{fig1}. The water surface is in the $x-y$ plane.
Governing equations for the motion of surface water waves in such
a system can be obtained by invoking the Newton's second law and
the conservation of mass by assuming that the water is
incompressible. While the detailed derivation has been given
elsewhere \cite{Ye}, here we just list the final equations.

The displacement of the water surface is denoted by $\eta(\vec{r},
t)$. Its Fourier transformation is \BE \eta(\vec{r}, t) =
\frac{1}{2\pi}\int_{-\infty}^\infty d\omega e^{-i\omega
t}\eta_\omega(\vec{r}).\EE The equation of motion for the Fourier
component $\eta_\omega$ is derived as\cite{Ye} \BE
\nabla\left(\frac{1}{k^2}\nabla\eta_\omega(\vec{r})\right) +
\eta_\omega(\vec{r}) = 0, \label{eq:finalb}\EE where $\nabla =
\partial_x \vec{e}_x +
\partial_y \vec{e}_y$, and the wavenumber $k$ satisfies \BE
\omega^2 = gk(\vec{r})\tanh(k(\vec{r})h(\vec{r})).
\label{eq:dispersion}\EE For a fixed frequency $\omega$, the
wavenumber varies as a function of the depth $h(\vec{r})$.

Although Eq.~(\ref{eq:finalb}) can be used for arbitrary
topographical bottoms, in this paper we focus on the cylindrical
steps depicted in Fig.~\ref{fig1}. Furthermore, we assume that all
the steps are identical. When there is a stimulating source, the
transmitted waves will be scattered repeatedly at the steps,
forming an orchestral pattern of multiple scattering. Such a
multiple scattering process can be solved for any arrangement of
the steps with help from the theory devised by
Twersky\cite{Twersky}. The wave transmission can be computed. In
the computation, the transmission is normalized such that it is
unity when there are no scatterers. The details have been
presented in \cite{Ye}.

When the steps are regularly placed to form periodic lattices, the
frequency bands will appear and can be determine as follows. By
Bloch's theorem\cite{Mermin}, the displacement field $\eta_\omega$
can be expressed in the following form \BE \eta_\omega(\vec{r}) =
e^{i\vec{K}\cdot\vec{r}} \sum_{\vec{G}} C_\omega(\vec{G},
\vec{K})e^{i\vec{G}\cdot\vec{r}},\label{eq:eta}\EE where $\vec{G}$
is the vector in the reciprocal lattice and $\vec{K}$ the Bloch
vector\cite{Mermin}. In this case, the wavenumber $k$ also varies
periodically and we have the following expression \BE
\frac{1}{k^2} = \sum_{\vec{G}}
A_\omega(\vec{G})e^{i\vec{G}\cdot\vec{r}}.\label{eq:h}\EE For a
fixed $\omega$, the coefficients $A_\omega$ are determined from
Eqs.~(\ref{eq:dispersion}) and (\ref{eq:h}).

Substituting Eqs.~(\ref{eq:eta}) and (\ref{eq:h}) into
Eq.~(\ref{eq:finalb}), we get \BE
\sum_{\vec{G}'}Q_{\vec{G},\vec{G}'}(\vec{K},\omega)C_\omega(\vec{G}',
\vec{K}) = 0, \label{eq:final03}\EE with
$$Q_{\vec{G},\vec{G}'}(\vec{K},\omega) =
[(\vec{G}+\vec{K})\cdot(\vec{G}'+\vec{K})]A_\omega(\vec{G}-\vec{G}')
- \delta_{\vec{G}, \vec{G}'}.$$ The dispersion relation connecting
$\vec{K}$ and $\omega$, {\it i.~e.} the frequency bands, is
therefore determined by the secular equation \BE \mbox{det}\left[
(\vec{G}+\vec{K})\cdot(\vec{G}'+\vec{K})]A_\omega(\vec{G}-\vec{G}')
- \delta_{\vec{G}, \vec{G}'}\right]_{\vec{G},\vec{G}'} = 0.
\label{eq:s} \EE Special care has to be taken in solving the above
secular equation, since the initial dispersion relation in
Eq.~(\ref{eq:dispersion}) is non-linear. We use an iterative
procedure to find the zero point or the fixed point for the
determinant. To gain confidence with the computation, we have
applied the numerical codes to three special cases for which the
solution is relatively easy to obtain: a flat bottom, steps in
shallow and deep waters. We found that the results match well the
expectations.

\begin{figure}[hbt]
\vspace{10pt} \epsfxsize=3in\epsffile{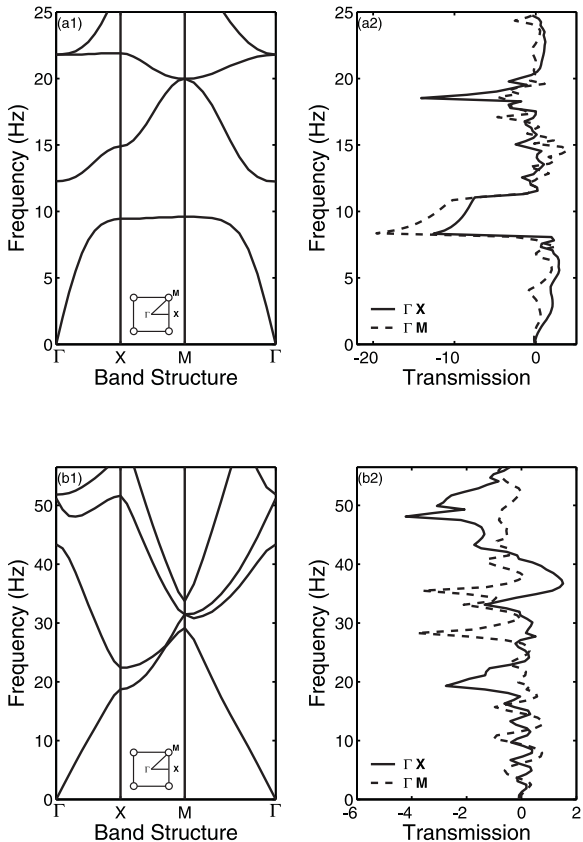} \caption{Right
Panel: Normalized transmission $\ln|T|^2|$ versus frequency for
square lattices cylindrical steps with two deferent heights. Left
panel: the calculated band structures for the corresponding
lattices. The inserted boxes in (a1) and (b1) denote the Brillouin
zone and illustrates the direction of wave transmission. For
example, $\Gamma X$ and $\Gamma M$ refer to [10] and [11]
direction respectively.}\label{fig2}
\end{figure}

First we consider the case of regular arrays. Figure \ref{fig2}
shows the results for the band structures and the transmission of
water waves across the arrays of cylindrical steps. With reference
to \cite{Nature}, the following parameters have been used in the
computation: lattice constant d = 2.5mm, cylinder radius a =
0.875mm, depth of the water h = 2.5mm. The heights for the steps
are 2.49 and 2.40 mm for (a) and (b) respectively. When computing
the transmission, a stimulating source is placed about one lattice
constant away from the arrays whereas the receiver is located at
about half of the lattice constant away on the other side of the
arrays. To ensure the stability of the results, enough modes and
number of steps have been considered. For instance, the maximum
mode number and the maximum array size considered are 9 and
30$\times$10 respectively. These numbers are considerably larger
than what has been computed previously.

Here it is shown that there is a complete band gap ranging from
9.6 to 12.3 Hz for the case in (a). The band structure calculation
in(a1) is fully supported by the independent transmission
calculation by the multiple scattering theory. Along the $\Gamma
X$ direction, the band structure shows that there is a partial
frequency gap from about 15 to 22 Hz. That is, waves whose
frequency lies within the range cannot propagate along this
direction. This partial gap also appears in the transmission
calculation shown by the solid line in (a2). However, the gap
depicted by the transmission calculation seems much narrower than
that obtained by the band structure calculation. We find that this
is due to the finiteness of the array. Reasons follows. Since a
point source was used to transmit waves, waves can be radiated to
various direction. Although the propagation along the $\Gamma X$
direction is prohibited in the presence of the partial gap, the
radiation into other directions may still have the chance to
arrive at the receiver, thus complicating the observation of the
partial gap.

We also find that the band structures and the transmission are
very sensitive to the arrangement and the height of the steps. As
an example, in Fig.~\ref{fig2}(b) we show the results for the same
lattice array as in (a) except that we change the height of the
steps from 2.49 mm to 2.40 mm. This slight change causes a
dramatic change in band structure. The complete band gap
disappeared. There are two partial gaps located at 20 and 45 Hz
respectively along the $\Gamma X$ direction. Within these two
gaps, the transmission is inhibited, as evidenced by the two
leftward valleys on the solid line in (b2). Compared to the
situation in (a), the transmission data match the band structures
better for the partial gaps.

We notice that there are two inhibited transmission valleys along
the $\Gamma M$ direction from the multiple scattering calculation,
referring to the dotted line in (b2). This phenomenon is
surprising, since in the frequency range concerned, roughly from
28 to 35 Hz, two frequency bands do show up in diagram (b1). A
possible explanation for this ambiguity may be that the two bands
are deaf. Such a deaf-band phenomenon has been recently observed,
for example, in acoustic systems\cite{Deaf} with a further support
from theoretical computations\cite{Ye2}.

\begin{figure}[hbt]
\vspace{10pt} \epsfxsize=3in\epsffile{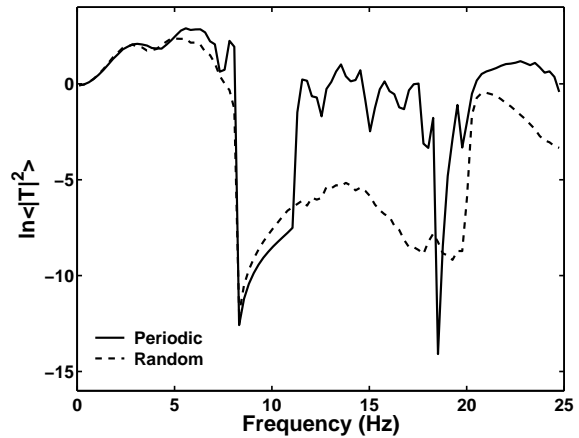} \caption{
Normalized transmission $\ln|T|^2$ versus frequency for periodic
and complete random arrays respectively. In the random case, the
transmission has been averaged over the random configurations.}
\label{fig3}
\end{figure}

Next we consider the effect of the randomization in the locations
of the steps. For brevity, we only show the result for the
complete random array. That is, the locations of the cylindrical
steps are completely random on the $x-y$ plane. The only
restraints are that no two steps should overlap with each other,
and the averaged distance between two nearest steps is kept as the
same as in the ordered case, i.~e. $d = 2.5$mm. The transmission
results are shown in Fig.~\ref{fig3}. The solid and dotted lines
separately refer to the transmission results for the propagation
along the [10] direction in the ordered case and for the complete
random case. At low frequencies, the disorder effect is not
obvious for the given sample size. In this regime, the scattering
by the steps is week. However, the transmission is significantly
reduced in the mid range of frequency. This observation is in
agreement with the case of acoustic scattering by arrays of rigid
cylinders located in air\cite{Ye}. For high frequencies, the
reduction due to the disorder is not as significant. This is
understandable. In the high frequency range, for example when $kh
>>1$, the effect of the steps on the wave propagation tends to
diminish. This can be seen from Eq.~(\ref{eq:dispersion}) which
reduces to $\omega^2 \approx gk$ in the high frequency regime.

In summary, we have considered water wave propagation over
bottom-mounted cylindrical steps. We found that complete band gaps
can appear in such a system in analogy with that in the photonic
or sonic crystals. The results also suggest that there might be
deaf-band phenomenon for the water waves.

{\bf Acknowledgements} The work received support from Fudan
University, NSC and NCU. One of us (YZ) is grateful to the
government of Shanghai for the Bai YuLan fund which makes his
visit to Fudan University possible. Useful help, discussion and
correspondence with Profs. J.-H. He, P. G. Luan, P. McIver, and J.
Zi are also thanked.

\end{document}